# Origin of the quasi-quantized Hall effect in $ZrTe_5$

S. Galeski[1]*, T. Ehmcke[2], R. Wawrzyńczak[1], P. M. Lozano[3], K. Cho[4], A. Sharma[4], S. Das[4], F. Küster[4], P. Sessi[4], M. Brando[1], R. Küchler[1], A. Markou[1], M. König[1], P. Swekis[1]. C. Felser[1], Y. Sassa[5], Q. Li[3], G. Gu[3], M. v. Zimmermann[7], O. Ivashko[7], D.I. Gorbunov[8], S. Zherlitsyn[8], T. Förster[8], S. S. P. Parkin[4], J. Wosnitza[8, 9], T. Meng[2], J Gooth[1,9#]

[1]*Max Planck Institute for Chemical Physics of Solids, Nöthnitzer Straße 40, 01187 Dresden, Germany.*

[2]*Institute for Theoretical Physics and Würzburg-Dresden Cluster of Excellence ct.qmat, Technische Universität Dresden, 01069 Dresden, Germany*

[3]*Condensed Matter Physics and Materials Science Department, Brookhaven National Laboratory, Upton, NY, USA.*

[4]*Max Planck Institute of Microstructure Physics, Weinberg 2, 06120 Halle (Saale), Germany.*

[5]*Department of Physics, Chalmers University of Technology, SE-412 96 Gothenburg, Sweden.*

[6]*Institut für Festkörper- und Materialphysik, Technische Universität Dresden, 01062 Dresden, Germany.*

[7]*Deutsches Elektronen-Synchrotron DESY, Notkestraße 85, D-22607 Hamburg, Germany*

[8]*Hochfeld-Magnetlabor Dresden (HLD-EMFL) and Würzburg-Dresden Cluster of Excellence ct.qmat, Helmholtz-Zentrum Dresden-Rossendorf, 01328 Dresden, Germany.*

[9]*Institut für Festkörper- und Materialphysik, Technische Universität Dresden, 01062 Dresden, Germany.*

*stanislaw.galeski@cpfs.mpg.de, # johannes.gooth@cpfs.mpg.de

## Abstract

**The quantum Hall effect (QHE) is traditionally considered to be a purely two-dimensional (2D) phenomenon. Recently, however, a three-dimensional (3D) version of the QHE was reported in the Dirac semimetal $ZrTe_5$. It was proposed that this arises from a magnetic-field-driven Fermi surface instability, transforming the original 3D electron system into a stack of 2D sheets. Here, we report thermodynamic, spectral, thermoelectric and charge transport measurements on such $ZrTe_5$ samples. The measured properties: magnetization, ultrasound propagation, scanning tunneling spectroscopy, and Raman spectroscopy, show no signatures of a Fermi surface instability, consistent with in-field single crystal X-ray diffraction. Instead, a direct comparison of the experimental data with linear response calculations based on an effective 3D Dirac Hamiltonian suggests that the quasi-quantization of the observed Hall response emerges from the interplay of the intrinsic properties of the $ZrTe_5$ electronic structure and its Dirac-type semi-metallic character.**

## Main text

Electrons subject to a magnetic field $B$, are forced to move on curved orbits with a discrete set of energy eigenvalues - the Landau levels (LLs). By increasing $B$, the LLs shift through the Fermi level $E_F$ one after the other, leading to quantum oscillations in transport and thermodynamic quantities.[1] At sufficiently large magnetic field, where only a few LLs are occupied, 2D electron systems (2DESs) enter the quantum Hall regime[2–5]. This regime is characterized by a fully gapped electronic spectrum in the bulk and current-carrying gapless edge states, leading to quantization of the Hall conductance $G_{xy} = \nu e^2/h$, where $\nu$ is the Landau level filling factor, $e$ is the elementary charge, and $h$ is Planck's constant. The situation is different in three dimensions. Instead of fully gapping the bulk of the 3D electron gas, high magnetic fields confine the electron motion in the plane perpendicular to the magnetic field still allowing them to freely move along the field, making the electron motion one-dimensional. Hence, current flow is allowed in the direction parallel to $B$. Crucially, the missing gap spoils quantization of $G_{xy}$. However, it is predicted that a 3D version of the QHE could occur in semi-metals and doped semiconductors,[6–9] in which the application of a magnetic field would lead to a Fermi surface instability, causing a periodic modulation of the electron density along the direction of $B$. Such modulation can effectively be thought of as a stack of 2DESs, each layer being in the quantum Hall regime. The signature of such a 3D quantum Hall system is that the Hall conductivity exhibits plateaus of $\sigma_{xy} = \nu e^2/h \cdot G_z/2\pi$ that are accompanied by a vanishing longitudinal electrical conductivity $\sigma_{xx}$.[6] $G_z$ is the reciprocal lattice vector of the modulation along $B$.

Recently, a quantized Hall response has been observed in the prototypical 3D Dirac-semimetal[10–13] materials $ZrTe_5$[14] and $HfTe_5$.[15,16] Particularly, the Hall resistivity $\rho_{xy}$ has been found to exhibit a plateau , scaling with $(e^2/h \cdot k_{F,b}/\pi)^{-1}$ for the magnetic field aligned with the crystal's $b$-axis when only the last LL is occupied (quantum limit). $k_{F,b}$ is the Fermi wave vector along the crystal $b$-axis at zero magnetic field. The observed scaling of the plateau height with $k_{F,b}/\pi$ has been interpreted as indicating a correlation-driven origin of the Hall effect. It was suggested[14] that the applied magnetic field leads to a Fermi surface instability and formation of a charge density wave (CDW) with a wavelength of $\lambda_z = \pi/k_{F,b}$ along the magnetic field. In a CDW, the density of the electrons and the position of the lattice atoms are periodically modulated with a wavelength, usually much larger than the original lattice constant.[17] CDWs usually the energetically preferred ground state of interacting quasi-one-dimensional conductors due to the almost perfect nesting of the Fermi surface.[17] In 3D systems, the dimensional reduction of the energy spectrum in high magnetic fields, in principle supports a scenario of a field induced CDW, in which each CDW modulation 'layer' contributes the conductance of one 2D quantum Hall system to the total bulk Hall conductivity.

The main argument for the CDW origin of the Hall plateaus in Ref. 15 is the vanishing longitudinal resistance in the quantum limit, which was interpreted as a consequence of a fully gapped electronic structure in the bulk and non-dissipative edge channels in each electron layer. However, the longitudinal resistivity seems to remain finite in the majority of the $ZrTe_5$[14] and $HfTe_5$[15,16] samples despite the presence of pronounced plateaus in the Hall resistivity (even down to 50 mK). Thus, it is desirable to investigate the physics of $ZrTe_5$ in magnetic fields beyond simple charge transport experiments.

In transport, a CDW transition typically manifests as an abrupt increase of electrical resistance due to the gapping of the Fermi surface and non-ohmic transport characteristics.[17] However, a CDW ground state cannot be identified by transport measurements alone, as such features also exist in materials without a CDW transition.[18–22] Both gapping of the Fermi surface and emergence of a periodic charge modulation should clearly impact thermodynamic, structural and spectral properties.[1] Gap openings in the electronic density of states (DOS) in non-magnetic materials can, for example, be probed by magnetization measurements,[23] thermoelectric coefficients[24] and scanning tunneling spectroscopy[17]. The corresponding periodic charge modulation typically leads to change in the phonon spectrum, which can be directly seen in Raman scattering[17], X-ray diffraction[25,26] and ultrasound propagation measurements[27,28]. In addition, Raman spectroscopy can directly probe the Raman active CDW amplitude mode[17].

In our study, we have used similar $ZrTe_5$ samples as those studied in Ref. [14], grown by tellurium flux method and applied all the after mentioned experimental techniques (see Methods for details). In total, we have investigated over 20 samples from which we exemplarily show data from Samples A, B, C, D, E, F, G and H, all exhibiting consistent results (Supplementary

Table S1). Upon cooling in $B = 0$ T, the longitudinal resistivity $\rho_{xx}$ increases with decreasing temperature $T$ until reaching a maximum at $T_L = 90$ K (Fig.1a, Supplementary Fig. S1 and Ref.[14]). This maximum has been previously observed in $ZrTe_5$ and has been attributed to a Lifshitz transition that is accompanied by a change of charge-carrier type.[29] Consistently, the sign of the zero-field Seebeck coefficient $S_{xx}$ changes sign at $T_L$,[11,30] indicating electron-type transport for $T < T_L$ and hole-type transport above.

Similarly to previous studies,[14,16] we find Shubnikov-de Haas oscillations in $\rho_{xx}(B)$ at low temperatures ($T < 30$ K) that are consistent with a single ellipsoidal Fermi (Fig. 1b). The magneto-transport measurement configuration is sketched in Fig. 1c. Further analysis of the Shubnikov-de Haas oscillations reveal that the quantum limit is reached at magnetic fields as low as $B_{QL} = 1.8$ T for the magnetic field applied along the $b$-axis (Supplementary Fig. S3, Supplementary Fig. S6, Supplementary Fig. S7 and Supplementary Fig. S14) and a relatively small effective mass of the order of 0.01 $m_0$ (Supplementary Table S1), which is in agreement with Dirac fermions ($m_0$ is the bare electron mass).[13] Importantly, in this field configuration (see for example Fig. 1b), all studied samples show a pronounced plateau in $\rho_{xy}(B)$ in the quantum limit with a height close to $(h/e^2)\ \pi/k_{F,b}$ as reported in Ref. [14]. However, $\rho_{xx}(B)$ does not vanish in any of the measured samples, even when cooled to 50 mK[16], consistent with Landau quantization of a 3D Fermi pocket. While $\rho_{xx}(B)$ remains finite in all samples, it is always much smaller than $\rho_{xy}(B)$ at low temperatures and thus the Hall conductivity $\sigma_{xy} = \rho_{xy}/(\rho_{xx}^2+\rho_{xy}^2)$ reduces to $\sigma_{xy} \approx 1/\rho_{xy}$, enabling the observation of the quantization in $\rho_{xy}$.

To test the hypothesis of a field-induced CDW, we have first investigated $S_{xx}(B)$ at low temperatures. $S_{xx}$ is particularly sensitive to changes in the electronic DOS, i.e. gapping of the Fermi surface, because it is proportional to the energy derivative of the DOS at $E_F$ (see SI and Ref [31] for a detailed discussion). However, we do not find signatures of a CDW in the $B$- dependent $S_{xx}$ with $B$ applied along the $a$, $b$ and $c$ crystallographic axes and for various angles in between them. Instead, the data shows quantum oscillations (Fig. 1 d-f and Supplementary Fig. S4 and Supplementary Fig. S9) that are consistent with a single ellipse-shaped 3D electron pocket at $E_F$ (Fig. 1 g-i) as observed in $\rho_{xx}(B)$.

Next, we have measured the magnetization of $ZrTe_5$ across the hypothetical phase boundary proposed in Ref. [14]. Magnetization is a thermodynamic quantity with the paramagnetic contribution directly proportional to the DOS and hence, highly sensitive to phase transitions,[23] such as for the formation of a CDW (see SI for a detailed discussion). Fig. 2a and b show the temperature dependence of magnetization measured at 2 T and 100 mT with the field applied along the $b$-axis – the field configuration in which the Hall plateaus are seen. The two investigated field values are chosen to cover two regimes: at 2T the system is in the quantum limit and at 100 mT multiple LLs are occupied. In both regimes, we find that the magnetization does not show any signatures of the formation of a CDW. Investigation of the magnetization as a function of magnetic field (Fig. 2 d and Supplementary Fig. S5) at low temperature reveals pronounced de Haas-van Alphen oscillations on top of a roughly linear background. The de Haas-van Alphen oscillations are in agreement with the Landau quantization of a single Fermi pocket and, hence, consistent with the observations in $\rho_{xx}(B)$ and $S_{xx}(B)$.[32] The strong linear diamagnetic background is indicative of a small effective mass, as expected for Dirac Fermions.[32,33]

Independently from transport and thermodynamic probes, the gap opening in the electronic structure, related to the formation of a CDW, could be probed using scanning tunneling spectroscopy (STS), since STS provides direct information about the DOS as a function of energy. Therefore, STS would directly corroborate the presence of a field induced gap tied to the formation of a CDW.[17] We have performed STS experiments on our $ZrTe_5$ samples at 0.4 K in fields up to 5 T applied along the crystallographic $b$-axis (Fig. 2e and Supplementary Fig.

S6). The measurements did not reveal any features in the density of states that could be attributed to a CDW formation. The DOS at $E_F$ remained finite even deep in the quantum limit, consistent with the observed Landau quantization of a 3D Fermi surface in $\rho_{xx}(B)$, $S_{xx}(B)$ and $M(B)$ and finite longitudinal resistivity. The observed band edges (labelled as $CB_1$, $CB_2$, and $VB_1$ in Fig. 2e) are consistent with previous experiments on $ZrTe_5$[34] and remain fixed with respect to each other – a characteristic feature of the lowest LLs of Dirac fermions.[35]

In addition, we have performed ultrasound propagation and attenuation experiments. Measurements of the sound velocity probe the system's elastic modulus, and thus a thermodynamic quantity that is very sensitive to sudden changes in the free energy, such as those caused by a CDW gap opening.[27,28,36] Ultrasound propagation measurements can provide information about the electron phonon coupling, a quantity crucial for the formation of a CDW (see Methods and SI for a detailed discussion). Fig. 2 f and g show the variation of the sound velocity $\Delta v_s/v_s$ and the sound attenuation $\Delta\alpha$ of the longitudinal sound modes (propagation along the $a$-axis and longitudinal polarization vector along the $a$-axis) as a function of magnetic field applied along the $b$-axis at 2 K. We find that both $\Delta v_s/v_s$ and $\Delta\alpha$ do not exhibit any anomalies that could indicate a phase transition. Instead, $\Delta v_s/v_s$ and $\Delta\alpha$ consistently reflect the quantum oscillations observed in $\rho_{xx}(B)$, $S_{xx}(B)$ and $M(B)$ that stand for a single Fermi pocket at $E_F$ (see also Supplementary Fig. S5 for the transverse mode) It has been pointed out[37] that for a CDW to emerge in the quantum limit regime above 1.5T, the strength of the effective electron-electron interaction generated from electron-phonon coupling should be around $g_0$ = 537.3 eV/nm. Our analysis of the quantum oscillations in $\Delta v_s/v_s$, however, suggests a much smaller coupling constant $g_0$ = 0.015 eV/nm (see Methods). Since $g_0$ depends quadratically on the electron-phonon coupling, we find that the electron-phonon-coupling in $ZrTe_5$ is too small by two orders of magnitude for a CDW to appear.

A field-induced phase transition into a CDW ground state is expected to lead to a modification of the crystal structure due to the new emerging charge modulation. To check if this is the case in $ZrTe_5$ we have performed X-ray diffraction measurements at zero field and in the quantum limit at 2 K. The formation of a CDW is expected to lead to the emergence of satellite superstructure peaks in the X-ray spectra with corresponding *k*-vectors of $G_z = 2k_{F,b}$ adjacent to the main Bragg reflections.[17] The results of *Q*-scans along the *b*-direction of the crystal in the vicinity of the (010)-Bragg peak with the field aligned along the *b*-axis are shown in Fig. 3 a. We find that the (010) Bragg peak does not change up to 2 T, and - most importantly - does not show any emerging satellite reflections.[26] The X-Ray experiments indicate that the zero-field crystal structure of $ZrTe_5$[38] is maintained in finite fields up to the quantum limit.

This result is confirmed by the magnetic field-dependent Raman spectroscopy measured at 2 K. The formation of CDW would cause additional peaks[17,39] to appear in the Raman spectra of $ZrTe_5$.[40,41] However, apart from small changes of the amplitude of the vibration modes, the Raman spectra do not change upon application of magnetic fields up to 9 T (Fig. 3b). In particular the existing phonon modes do not show change in Raman shift and no new spectral features appear that could be related to absorption by the Raman active CDW amplitude mode. Our experimental investigations indicate that the state underlying the emergence of Hall plateaus in the quantum limit of $ZrTe_5$ is in fact gapless and exhibits the behavior of a Fermi sea of a Landau-quantized 3D Dirac semimetal.

To directly compare the effect of Landau quantization of a 3D Dirac system on the Hall resistivity, we performed linear response calculations of the electrical transport properties, based on an anisotropic Dirac Hamiltonian, with a magnetic field along the *z*-direction (see Methods). As a cross check, we also compute $S_{xx}$,, $S_{xy}$, $\Delta v_s/v_s$, and *M* (see Methods for details), using the same model. Although the inverse Landau level broadening and the transport relaxation time are in general sensitive to different scattering mechanisms, the calculations were performed assuming that both quantities are the same. This approximation has the advantages of being transparent to interpretation and that the calculated quantities can be directly related to the band structure. However, as a side effect, the calculated $\rho_{xx}$ is slightly different than seen in the experiment (Fig. 4). More importantly, our model reproduces all other experimental data even on a quantitative level (Fig. 2 and Fig. 4) and explains the key observations of the electrical transport in $ZrTe_5$: First, at high magnetic fields, Landau quantization leads to plateaus in $\rho_{xy}(B)$ with a height of $(h/e^2)\ \pi/k_{F,b}$ in the quantum limit. For a 3D system, the Landau levels form continuous Landau bands dispersing parallel to the applied field. The Hall conductivity is the sum of conductance quanta over all occupied Landau bands $\nu$ and wave numbers $k_b$, $\sigma_{xy}(B) = \sum_{\nu\ occ.} \int_{k_b\ occ.} \frac{dk_b}{2\pi} \frac{e^2}{h} = \sum_{\nu\ occ.} \frac{2k_{F,b,\nu}(B)}{2\pi} \frac{e^2}{h}$, where $k_{F,\nu,b}(B)$ is the Fermi wave number parallel to the field. While $k_{F,b,\nu}$ in general depends on $B$, a characteristic feature of Dirac systems is that the lowest LL does not shift with magnetic field[35] (see Supplementary Fig. S16) and $\sigma_{xy}(B)$ becomes $\frac{2k_{F,b}}{2\pi} \frac{e^2}{h}$ in the quantum limit. Second, while $\rho_{xx}(B)$ shows Shubnikov-de Haas oscillations with minima at the center of the Hall plateaus, it is *finite* for all magnetic fields, due to the gapless Landau band structure.

Magnetotransport in 2DESs with localized states is often described by effectively fixed chemical potentials, which leads to well-defined quantum Hall plateaus. In contrast, the large carrier densities $n$ in ordinary 3D metals, such as copper, usually force the particle number to be conserved in order to avoid large charging energies. This implies the chemical potential to vary as a function of magnetic field. In materials with small Fermi surfaces, such as Dirac semimetals, the charging energies on the contrary remain relatively small, since the absolute change in $n$ with $B$ at fixed $E_F$ remains small. Defects can furthermore induce localized states that absorb some of the conduction electrons,[42] and thermally activated higher bands may similarly serve as reservoirs of states that do not contribute to transport.[43] As a result, the charge carrier density in the conduction band can vary without the sacrifice of overall charge neutrality. In fact, the measured STS (Fig 2e) shows that $E_F$ does not shift with respect to the edge of the conduction band $CB_1$, i.e. lowest LL, up to 5 T. A variable particle number in the conduction band furthermore agrees with the observed Hall response that does not show the smooth behavior $\rho_{xy} = B/ne$ (which would be expected if particle number were fixed). We find that the behavior of $ZrTe_5$, up to fields of several Tesla is in fact well-described by a model including only the itinerant electrons, but keeping $E_F$ fixed.

We now address the expected generality of our results. Our model provides a way to derive and understand the Hall conductivity of a genuine 3D electron systems from the conductance quantum scaled by a characteristic length, when only the lowest few Landau bands are occupied. This length scale, however, does not relate to a 2D spatial confinement, but rather to an intrinsic momentum vector of the 3D electron system, which is given by the electronic band structure - a result that is also obtained for a stack of coupled quantum Hall layers.[16] Therefore, we expect quasi-quantized features in the Hall conductivity to be observed in the quantum limit of generic 3D metals and semimetals, regardless of the precise band structure, Fermi level, or purity as long as the particle number of conduction band electrons is not strictly preserved. However, the shape of the quasi-quantized features in $\sigma_{xy}$ will depend on fine details of the 3D Landau level spectrum. Our findings promise to explain the often puzzling plateaus appearing in Hall measurements in many other low charge carrier semimetals such as NbP[44] or HgSe[45] and in slightly doped semiconductors such as InAs[42,46] and InSb[47,48], to name a few of them.

In summary, we have shown that $ZrTe_5$ exhibits quasi-quantized plateaus in the Hall resistivity that scale with $\frac{2\pi}{2k_{F,b}}\frac{h}{e^2}$ in the quantum limit due to an interplay of its Dirac band structure and low charge carrier density. Our model can be directly applied to Landau quantization of generic 3D band structures, and also to the 3D anomalous Hall and Spin-Hall systems. Our findings establish the Hall effect in $ZrTe_5$ as a truly three-dimensional relative of the quantum Hall effect in 2D systems, and a prime candidate for the observation relativistic chiral surface states[18, 19].

## Methods

*Sample synthesis and preparation*

High-quality single-crystal $ZrTe_5$ samples were synthesized with high-purity elements (99.9999% zirconium and 99.9999% tellurium), the needle-shaped crystals (about $(0.1 \times 0.3 \times 20)$ mm$^3$) were obtained by the tellurium flux method. The lattice parameters of the crystals were structurally confirmed by single crystal X-ray diffraction. Prior to transport measurements Pt contacts were sputter deposited on the sample surface to ensure low contact resistance. The contact geometry was defined using Al hard masks. Prior to Pt deposition the sample surfaces were Argon etched and a 20 nm Ti buffer layer was deposited to ensure good adhesion of the contacts. Deposition was conducted using a BESTEC UHV sputtering system. This procedure allowed us to achieve contact resistance of the order of 1-2 Ohm.

*Sample environment*

The pulsed magnetic field experiments up to 70 T were carried out at the Dresden High Magnetic Field Laboratory (HLD) at HZDR, a member of the European Magnetic Field Laboratory (EMFL). All transport measurements up to ±9 T were performed in a temperature-variable cryostat (PPMS Dynacool, Quantum Design), equipped with a dilution refrigerator insert and a horizontal rotator.

*Electrical and thermoelectric transport Measurements*

To avoid contact-resistance, only four-terminal measurements were carried out. The longitudinal $\rho_{xx}$ and Hall resistivity $\rho_{xy}$ were measured in a Hall-bar geometry with standard lock-in technique (Zurich instruments MFLI and Stanford Research SR 830), with a frequency selected to ensure a phase shift below 1 degree – typically in the range of $f$ = 10-1000 Hz across a 100 kΩ shunt resistor. In addition, some samples were measured, using a Keithley Delta-mode resistance measurement setup for comparison. In both measurement modes the electrical current was always applied along the *a*-axis of the crystal and never exceeded 10 μA in order to avoid self-heating. Thermoelectric measurements were performed using the same electrical contacts as for electrical transport measurements. In order to supply a substantial temperature gradient across the sample despite its high thermal conductivity the sample was semi suspended with a heater attached to the free hanging end. In addition, a set of two Cernox Cx-1060 thermometers were attached to the sample in order to measure the temperature gradient. In order to obtain thermoelectric data within the linear response regime the applied gradient was kept at less than 10% of the sample temperature. The thermal voltage was measured using a Keithley 2182A nanovoltmeter. In order to avoid the influence of parasitic thermal voltages on the cryostat cables a background measurement without applied power was carried out both for Seebeck and Nernst measurements at all measured fields and temperatures.

*Ultrasound propagation measurements*

Ultrasound measurements in pulsed magnetic fields up to 10 Tesla were performed using a phase-sensitive pulse-echo technique. Two piezoelectric lithium niobate ($LiNbO_3$) resonance transducers were glued to opposite parallel surfaces of the sample to excite and detect acoustic waves. The transducer surfaces were polished using a focused Ion beam in order to ensure that the transducer attachment surfaces were smooth and parallel. The longitudinal and transverse acoustic waves with were propagated along the *a*-axis with the transverse polarization vector along the *c*-axis. Relative sound-velocity changes $\Delta v/v$, and sound attenuation $\Delta\alpha$, were measured for field applied along the b axis. The longitudinal and transverse ultrasound propagation were measured at 28 MHz and 313 MHz respectively.

*Magnetization measurements*

Magnetization measurements were conducted in a standard Quantum Design VSM MPMS equipped with a 7 Tesla superconducting magnet. For measurements the samples were attached on quartz sample holders and glued using a small amount of GE-varnish. In order to avoid parasitic contributions in magnetic measurements at small fields where the magnetic response of $ZrTe_5$ is small the background magnetization of the quartz holder together with the adhesive was measured and subtracted from the data.

*In-field single crystal X-ray diffraction*

In-field single crystal X-ray diffraction measurements have been performed at the Petra III P21.1 beamline at DESY (Hamburg, Germany). Measurements were performed in standard cryostat equipped with a 10 Tesla horizontal superconducting magnet. For the measurements the sample *b*-axis was aligned along the field direction, *c*-axis was in the scattering plane while a-axis was vertical. In order to detect new satellite peaks, several reciprocal-space directions were scanned both at 0 and 2 Tesla. Namely, we have performed *k*-scans in the low-*Q* range [(0,0,0) – (0,4,0)] and, at high-Q range [(0,12,*l*) – (0,16,*l*) with $l = 0$, 0.5, 0.75, 1 and (0,16,l) – (0,20,l) for $l = 0$, 1] where the x-ray structure factor is expected to be stronger, For background free measurements we have used a CdTe Amptek point detector with a combination of a Ge-gradient Si analyzer and a 101.7 keV incident energy beam.

*Scanning tunneling microscopy*

$ZrTe_5$ crystals were cleaved in ultra-high vacuum (base pressure ~ $1\times10^{-10}$ mbar) at room temperature and immediately inserted into a cryogenic scanning tunneling microscope operated at $T = 1.9$ K. The relatively weak van der Waals bonding between the $ZrTe_5$ layers offers a natural cleaving plane, resulting in a Te-terminated surface. Scanning tunneling microscopy measurements were acquired using electrochemically etched W tips. Spectroscopic data were measured using a lock-in technique, modulating the bias with 1 mV (r.m.s.) at 733 Hz.

*Theory - the Dirac Hamiltonian*

*Ab-initio* calculations[10] suggest $ZrTe_5$ to be a Dirac semimetal whose low-energy band structure at zero field can be modelled using an anisotropic Dirac Hamiltonian:[14]

$$H(\boldsymbol{k}) = m\tau_3\sigma_0 + \hbar(v_a k_a \tau_1\sigma_3 + v_c k_c \tau_2\sigma_0 + v_b k_b \tau_1\sigma_1),$$

with $\tau_i$ ($\sigma_i$) denoting Pauli matrices acting on the orbital (spin) degree of freedom and $k_j$ and $v_j$ denote the components of the momentum vector $\boldsymbol{k}$ and Fermi velocity in the $j$-th direction, respectively. $m$ accounts for the zero-field-gap. The magnetic field $\boldsymbol{B} = \mathrm{rot}\,\boldsymbol{A}$ enters the Hamiltonian through the orbital effect which is implemented via the usual substitution: $\hbar\boldsymbol{k} \to \hbar\boldsymbol{k} + e\boldsymbol{A}$, and the Zeeman effect is introduced via $H_Z = -\frac{1}{2} g\mu_B \tau_0 \boldsymbol{\sigma} \cdot \boldsymbol{B}$**.** Here, $\boldsymbol{A}$ is the vector potential, $g$ is the Landé g-factor and $\mu_B$ is the Bohr magneton. The spectral function for the corresponding $\nu$-th Landau band at b wave number $k_b$ is:

$$\rho_{\nu,k_b}(\epsilon) = -\frac{1}{\pi}\mathrm{Im}\left(\epsilon - E_\nu(k_b) + i\frac{\hbar}{2\tau_Q}\right)^{-1},$$

with the quantum lifetime $\tau_Q$. For the calculation of the electric- and thermoelectric response functions, we have used the parameters: $m = 10$ meV, $g = 10$, $v_a = 116392\,\frac{m}{s}$, $v_b = 15340\,\frac{m}{s}$, $v_c = 348875\,\frac{m}{s}$, $\mu = 12.745$ meV, $\tau_Q = 3.064$ ps

*Linear response theory – electric and thermoelectric transport*

Electric transport is calculated in linear response via the Kubo formula. The longitudinal and Hall conductivities are obtained from the respective current-current correlation functions. The Green's functions entering these equations contain an imaginary self-energy to account for the scattering-induced lifetime of quasiparticles. The thermoelectric conductivity tensor $\hat{\epsilon}$ is obtained from the zero-temperature electrical conductivity tensor $\hat{\sigma}$ via integration as $\hat{\epsilon} = -\frac{1}{|e|T}\int_{-\infty}^{\infty} \mathrm{d}\epsilon\,(\epsilon - \mu)\left(-\frac{\partial n_F(\epsilon-\mu)}{\partial\epsilon}\right)\hat{\sigma}(T = 0, \mu = \epsilon)$. The thermopower tensor $\hat{S}$, which contains the Seebeck and Nernst coefficients, is calculated via $\hat{S} = \hat{\sigma}^{-1}\hat{\epsilon}$.

*Partition function theory - magnetization*

The magnetization $M$ is calculated as the derivative of the free energy $F$ with respect to the magnetic field, $M = -\frac{1}{V}\frac{dF}{dB}$, where $V$ is the volume. The free energy is in turn defined via the canonical partition sum. Here have neglected Landau level broadening ($\tau_Q \to \infty$), but have otherwise used the same parameters as described above (see Theory - the Dirac Hamiltonian).

*Theoretical model for phonon velocity renormalization*

We consider a longitudinal acoustical phonon that propagates along the magnetic field direction (parallel to the $b$-axis). In the long wavelength limit the free phonon energy is given by $\hbar\Omega_{\boldsymbol{q}} = v_s q$, where $v_s$ is the sound velocity and $q$ is the phonon momentum. Electron-phonon interactions induce a self-energy contribution $\Pi^{\mathrm{R}}(\boldsymbol{q}, \omega = \Omega_{\boldsymbol{q}})$ to the retarded phonon propagator. Its real part corresponds to the energy correction due to the electron-phonon interactions, $\mathrm{Re}\{\Pi^{\mathrm{R}}(\boldsymbol{q}, \omega = \Omega_{\boldsymbol{q}})\} = \hbar\Delta\Omega_{\boldsymbol{q}} = \Delta v_s q$.[49] Evaluating the self-energy in a random-phase approximation at $T = 0$ and for finite Landau level broadening $\Gamma > 0$ yields:

$$\frac{\Delta v_s}{v_s} = \lambda B \sum_{\nu} \int_0^{\infty} \mathrm{d}k_b \; \rho_{\nu,k_b}(\mu) \, ,$$

where $\rho_{\nu,k_b}$ is the spectral function (see section 1.3) of the $\nu$-th Landau band at b wave number $k_b$ and $\lambda = -D_1^2|e|/(4\pi^2 \rho v_s^2 \hbar) = -1.139 \times 10^{-35} \frac{Jm}{T}$ is a fitting parameter. To account for sample-specific differences in the electronic band structure, we have chosen slightly different parameters in the effective Dirac Hamiltonian. Namely, we have changed the mass parameter to $m = 9.5$ meV, the g-factor to $g = 7$, the Fermi velocity $b$-component to $v_b$=16486 m/s and the quantum lifetime to $\tau_Q = 0.658$ ps in the calculation of the phonon velocity renormalization. The parameter $\lambda$ is used to approximate the effective electron-electron interaction strength $g_0 = -4\pi^2 \lambda v_s \hbar^2 \left( \left(2k_{F,b}\right)^2 + \kappa^2 \right)^2 / |e|$ for electrons at the Fermi level, $k_z = \pm k_{F,b}$, from the very

recent work of Qin et al.[37]. Here, $\kappa = \sqrt{|e|^3 B/(4\pi^2 \epsilon \hbar^2 v_b)}$ is the inverse Coulomb screening length with the dielelectric constant $\epsilon/\epsilon_0 = 25.3$ evaluated at $B = 1$ T.

## References

1. Landau, L. D. & Lifshitz, E. M. *Quantum mechanics: non-relativistic theory*. **3**, (Elsevier, 2013).

2. Klitzing, K. V., Dorda, G. & Pepper, M. New method for high-accuracy determination of the fine-structure constant based on quantized Hall resistance. *Phys. Rev. Lett.* **45**, 494–497 (1980).

3. Tsui, D. C., Stormer, H. L. & Gossard, A. C. Two-dimensional magnetotransport in the extreme quantum limit. *Phys. Rev. Lett.* **48**, 1559 (1982).

4. Yoshioka, D. *The quantum Hall effect*. **133**, (Springer Science & Business Media, 2013).

5. Halperin, B. I. Theory of the quantized Hall conductance. *Helv.Phys.Acta* **56**, 75–102 (1983).

6. Halperin, B. I. Possible states for a three-dimensional electron gas in a strong magnetic field. *Jpn. J. Appl. Phys.* **26**, 1913 (1987).

7. Kohmoto, M., Halperin, B. I. & Wu, Y.-S. Diophantine equation for the three-dimensional quantum Hall effect. *Phys. Rev. B* **45**, 13488 (1992).

8. Bernevig, B. A., Hughes, T. L., Raghu, S. & Arovas, D. P. Theory of the three-dimensional quantum Hall effect in graphite. *Phys. Rev. Lett.* **99**, 146804 (2007).

9. Koshino, M. & Aoki, H. Integer quantum Hall effect in isotropic three-dimensional crystals. *Phys. Rev. B* **67**, 195336 (2003).

10. Weng, H., Dai, X. & Fang, Z. Transition-metal pentatelluride $ZrTe_5$ and $HfTe_5$: A paradigm for large-gap quantum spin Hall insulators. *Phys. Rev. X* **4**, 11002 (2014).

11. Li, Q. *et al.* Chiral magnetic effect in $ZrTe_5$. *Nat. Phys.* **12**, 550 (2016).

12. Liang, T. *et al.* Anomalous Hall effect in $ZrTe_5$. *Nat. Phys.* **14**, 451–455 (2018).

13. Zheng, G. *et al.* Transport evidence for the three-dimensional Dirac semimetal phase in

ZrT e 5. *Phys. Rev. B* **93**, 115414 (2016).

14. Tang, F. *et al.* Three-dimensional quantum Hall effect and metal–insulator transition in ZrTe5. *Nature* **569**, 537–541 (2019).

15. Wang, P. *et al.* Approaching three-dimensional quantum Hall effect in bulk $HfTe_5$. *Phys. Rev. B* **101**, 161201 (2020).

16. Galeski, S. *et al.* Unconventional Hall response in the quantum limit of $HfTe_5$. *Nat. Commun.* **11**, 1–8 (2020).

17. Grüner, G. The dynamics of charge-density waves. *Rev. Mod. Phys.* **60**, 1129–1181 (1988).

18. Kawamura, M., Endo, A., Katsumoto, S. & Iye, Y. Non-ohmic vertical transport in multilayered quantum Hall systems. *Phys. E Low-dimensional Syst. Nanostructures* **6**, 698–701 (2000).

19. Druist, D. P., Turley, P. J., Maranowski, K. D., Gwinn, E. G. & Gossard, A. C. Observation of chiral surface states in the integer quantum Hall Effect. *Phys. Rev. Lett.* **80**, 365 (1998).

20. Pusep, Y. A., Guimarães, F. E. G., Arakaki, A. H. & de Souza, C. A. Spectroscopic evidence of extended states in the quantized Hall phase of weakly coupled GaAs/AlGaAs multilayers. *J. Appl. Phys.* **104**, 63702 (2008).

21. Pollak, M. & Shklovskii, B. *Hopping transport in solids*. (Elsevier, 1991).

22. Kittel, C. *Introduction to solid state physics*. **8**, (Wiley New York, 1976).

23. Mahan, G. D. *Many-particle physics*. (Springer Science & Business Media, 2013).

24. Zhu, Z., Yang, H., Fauqué, B., Kopelevich, Y. & Behnia, K. Nernst effect and dimensionality in the quantum limit. *Nat. Phys.* **6**, 26–29 (2010).

25. Gerber, S. *et al.* Three-dimensional charge density wave order in $YBa_2Cu_3O_6$ at high magnetic fields. *Science* **350**, 949–952 (2015).

26. Hoshino, H. F. and M. S. and S. X-ray diffraction study of the quasi-one-dimensional conductors (MSe 4 ) 2 I (M=Ta and Nb). *J. Phys. C Solid State Phys.* **18**, 1105 (1985).

27. Saint-Paul, M. *et al.* Elastic anomalies at the charge density wave transition in $TbTe_3$. *Solid State Commun.* **233**, 24–29 (2016).

28. Jericho, M. H., Simpson, A. M. & Frindt, R. F. Velocity of ultrasonic waves in 2 H-Nb $Se_2$, 2 H-$TaS_2$, and 1 T-$TaS_2$. *Phys. Rev. B* **22**, 4907 (1980).

29. Zhang, Y. *et al.* Electronic evidence of temperature-induced Lifshitz transition and topological nature in $ZrTe_5$. *Nat. Commun.* **8**, 1–9 (2017).

30. Jones, T. E., Fuller, W. W., Wieting, T. J. & Levy, F. Thermoelectric power of $HfTe_5$ and $ZrTe_5$. *Solid State Commun.* **42**, 793–798 (1982).

31. Behnia, K. & Aubin, H. Nernst effect in metals and superconductors: a review of concepts and experiments. *Reports Prog. Phys.* **79**, 46502 (2016).

32. Zhang, C.-L. *et al.* Non-saturating quantum magnetization in Weyl semimetal TaAs. *Nat. Commun.* **10**, 1–7 (2019).

33. Koshino, M. & Ando, T. Anomalous orbital magnetism in Dirac-electron systems: Role of pseudospin paramagnetism. *Phys. Rev. B* **81**, 195431 (2010).

34. Li, X.-B. *et al.* Experimental observation of topological edge states at the surface step edge of the topological insulator $ZrTe_5$. *Phys. Rev. Lett.* **116**, 176803 (2016).

35. Chen, R. Y. *et al.* Magnetoinfrared spectroscopy of Landau levels and Zeeman splitting of three-dimensional massless Dirac fermions in $ZrTe_5$. *Phys. Rev. Lett.* **115**, 176404 (2015).

36. Kittel, C. & Fong, C. *Quantum theory of solids*. **5**, (Wiley New York, 1963).

37. Qin, F. *et al.* Theory for the charge-density-wave mechanism of 3D quantum Hall effect. *Phys. Rev. Lett.* **125**, 206601 (2020).

38. Skelton, E. F. *et al.* Giant resistivity and X-ray diffraction anomalies in low-

dimensional $ZrTe_5$ and $HfTe_5$. *Solid State Commun.* **42**, 1–3 (1982).

39. Tsang, J. C., Smith Jr, J. E. & Shafer, M. W. Raman spectroscopy of soft modes at the charge-density-wave phase transition in 2 H− $NbSe_2$. *Phys. Rev. Lett.* **37**, 1407 (1976).

40. Zhou, Y. *et al.* Pressure-induced superconductivity in a three-dimensional topological material $ZrTe_5$. *Proc. Natl. Acad. Sci.* **113**, 2904–2909 (2016).

41. Qiu, G. *et al.* Observation of optical and electrical in-plane anisotropy in high-mobility few-layer $ZrTe_5$. *Nano Lett.* **16**, 7364–7369 (2016).

42. Haude, D., Morgenstern, M., Meinel, I. & Wiesendanger, R. Local density of states of a three-dimensional conductor in the extreme quantum limit. *Phys. Rev. Lett.* **86**, 1582 (2001).

43. Wang, C. Thermodynamically Induced Transport Anomaly in Dilute Metals $ZrTe_5$/$HfTe_5$. *arXiv Prepr. arXiv2011.01952* (2020).

44. Shekhar, C. *et al.* Extremely large magnetoresistance and ultrahigh mobility in the topological Weyl semimetal candidate NbP. *Nat. Phys.* **11**, 1–6 (2015).

45. Lonchakov, A. T., Bobin, S. B., Deryushkin, V. V & Neverov, V. N. Observation of quantum topological Hall effect in the Weyl semimetal candidate HgSe. *J. Phys. Condens. Matter* **31**, 405706 (2019).

46. Murzin, S. S. Metal-Hall insulator transition. *JETP Lett.* **44**, 56–59 (1986).

47. Egorov, F. A. & Murzin, S. S. Temperature dependences of the conductivity of n-type lnSb and n-type lnAs in the extreme quantum limit. *Zh. Eksp. Teor. Fiz* **94**, 15–320 (1988).

48. Murzin, S. S., Jansen, A. G. M. & Zeitler, U. Magnetoconductivity of metallic InSb in the extreme quantum limit. *Phys. B Condens. Matter* **165**, 301–302 (1990).

49. Zhang, S.-B. & Zhou, J. Quantum oscillations in acoustic phonons in Weyl semimetals. *Phys. Rev. B* **101**, 85202 (2020).

## Notes

The authors declare no competing financial interests.

## Author contributions

S.G and J.G. conceived the experiment. The single crystals were grown by G.G. The basic structural and transport properties of bulk crystals were measured and studied by P.M.L. and Q. L. S.G., R.W., A.M. and C.F. fabricated the final transport devices. S.G. and R.W. performed the transport experiments. S.G., M.K., D.G., S.Z., J.W. performed ultrasound propagation measurements. S. D., F. K. and P.S. performed the Scanning Probe Spectroscopy. K. C. and A. S. carried out the Raman experiments. S.S.P.P. supervised the Scanning Probe Spectroscopy and the Raman experiments. S.G., T.F., performed high field transport measurements. S.G, Y.S, M.Z, O.I prepared and executed the in-field X-ray diffraction experiments. S.G., M.B., P.S. and R.K. carried out the magnetization measurements. T.E. and T.M. provided the theoretical model of the three-dimensional quantum Hall effect. S.G., T.M., T.F. and J.G. analyzed the data. All authors contributed to the interpretation of the data and to the writing of the manuscript.

## Acknowledgments

T.M. acknowledges funding by the Deutsche Forschungsgemeinschaft via the Emmy Noether Programme ME4844/1- 1, the Collaborative Research Center SFB 1143, project A04, and the Cluster of Excellence on Complexity and Topology in Quantum Matter ct.qmat (EXC 2147). C.F. acknowledges the research grant DFG-RSF (NI616 22/1): Contribution of topological states to the thermoelectric properties of Weyl semimetals and SFB 1143. P.M.L, G.G. and Q.L. were supported by the US Department of Energy, Office of Basic Energy Science, Materials

Sciences and Engineering Division, under contract DE-SC0012704. J.W. acknowledges support from the DFG through the Würzburg-Dresden Cluster of Excellence on Complexity and Topology in Quantum Matter ct:qmat (EXC 2147, project-id 39085490), the ANR-DFG grant Fermi-NESt, and by Hochfeld-Magnetlabor Dresden (HLD) at HZDR, member of the European Magnetic Field Laboratory (EMFL). J.G. acknowledges support from the European Union's Horizon 2020 research and innovation program under Grant Agreement ID 829044 "SCHINES". We acknowledge DESY (Hamburg, Germany), a member of the Helmholtz Association HGF, for the provision of experimental facilities. Parts of this research were carried out at beamline P21.1 at PETRA III. Y.S. acknowledges funding from the Swedish Research Council (VR) with a Starting Grant (Dnr. 2017-05078) as well as Chalmers Area Of Advance-Materials Science.

## Supplementary Information

The Supplementary Information contains Supplementary Section S1 "Charge-carrier density and mobility from Hall measurements", Supplementary Section S2 "Mapping of the Fermi surface by analyzing Shubnikov-de Haas oscillations", Supplementary Section S3 "Calculation of the Hall conductivity tensor element", Supplementary Section S4 "Non-linear electrical transport in multilayer quantum Hall systems", Supplementary Section S5 "Derivative relations between electrical and thermoelectrical quantum transport coefficients in $ZrTe_5$", Supplementary Section S6 "Discussion on the sensitivity of our experiments for probing the hypothetical charge density wave in $ZrTe_5$", Supplementary Fig. S1 – S14, Supplementary Table S1 and Supplementary References.

## MAIN FIGURES

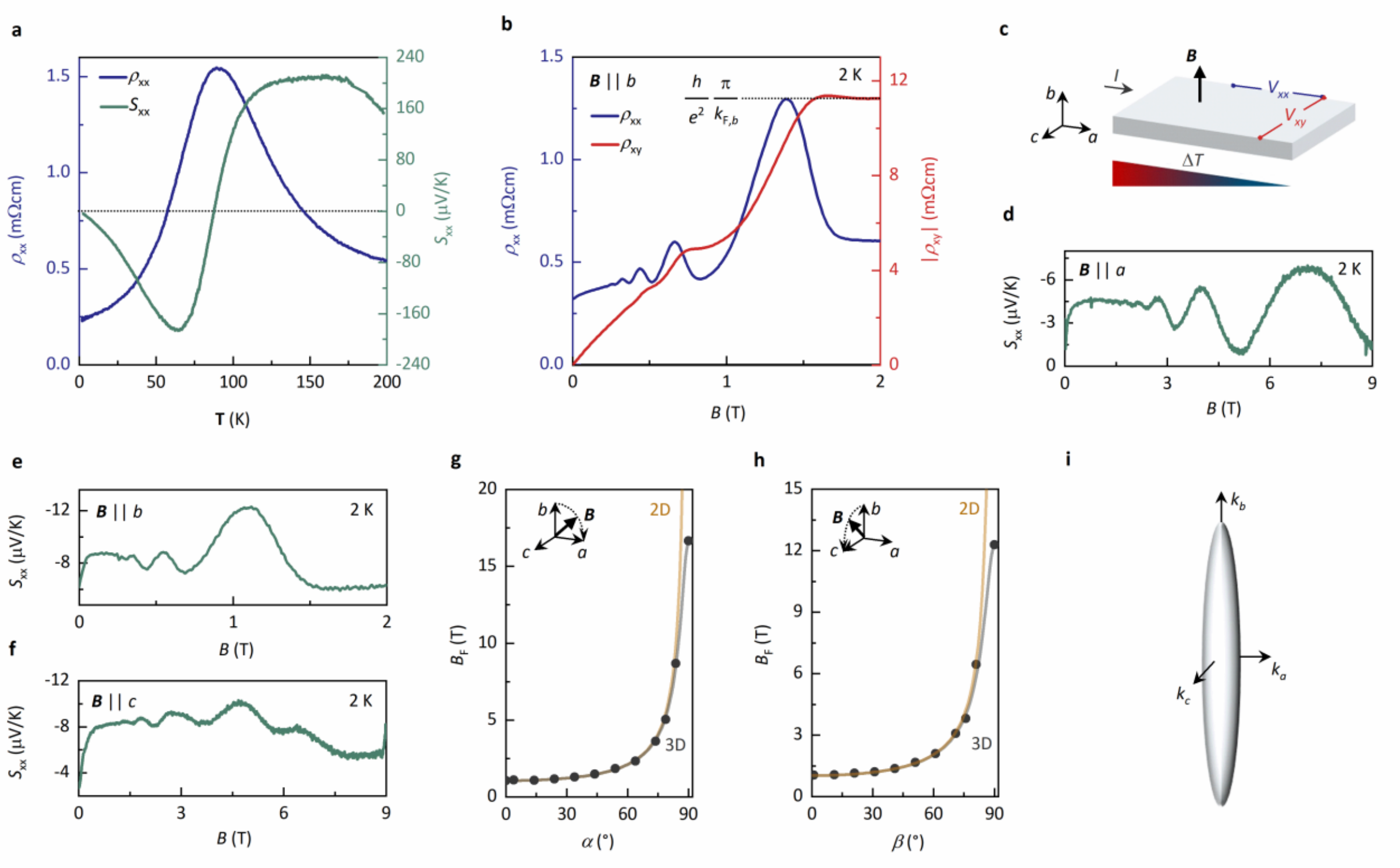

**Figure 1 | Three-dimensional morphology of the Fermi surface and quasi-quantized Hall effect in $ZrTe_5$. a**, Longitudinal electrical resistivity $\rho_{xx}$ and Seebeck coefficient $S_{xx}$ of Sample A as a function of temperature $T$ at zero magnetic field. **b,** $\rho_{xx}$ and Hall resistivity $\rho_{xy}$ as a function of $B$ applied along the $b$-axis at 2K, obtained on Sample B. The last Hall plateau scales with the experimentally extracted Fermi wave vector $k_{F,,b}$ along $B$ (Supplementary Table 1), the electron charge $e$ and the Planck constant $h$. **c**, Sketch of the transport measurement configurations with respect to the three crystal axes $a$, $b$, and $c$. The electrical current $I$ and the temperature gradient $\Delta T$ are applied along the $a$-axis. The corresponding longitudinal ($V_{xx}$) and Hall ($V_{xy}$) voltage responses are measured along the $a$-axis ($V$xx) and along $c$-axis, respectively. **d**, $S_{xx}$ as a function of $\boldsymbol{B}$ at 2 K with $\boldsymbol{B}$ applied along the *a,* **e**, along the $b$ and **f**, along the $c$-axis, measured on Sample A. **g**, Shubnikov-de Haas frequency $B_F$ as a function of angle ($\alpha$ and $\beta$) between $B$ and the $b$-axis, rotated within the $a$-$b$ plane and **h**, within the $b$-$c$ plane of Sample A.

The black dots represent the measurement data. The yellow lines represent fitting curves of a planar 2D Fermi surface-model to the data. The black lines represent fitting curves of an ellipsoidal 3D Fermi surface-model to the data. **i**, Sketch of the experimentally extracted Fermi surface of $ZrTe_5$ along the momentum vectors $k_a$, $k_b$ and $k_c$ in $a$, $b$, and $c$ direction, respectively.

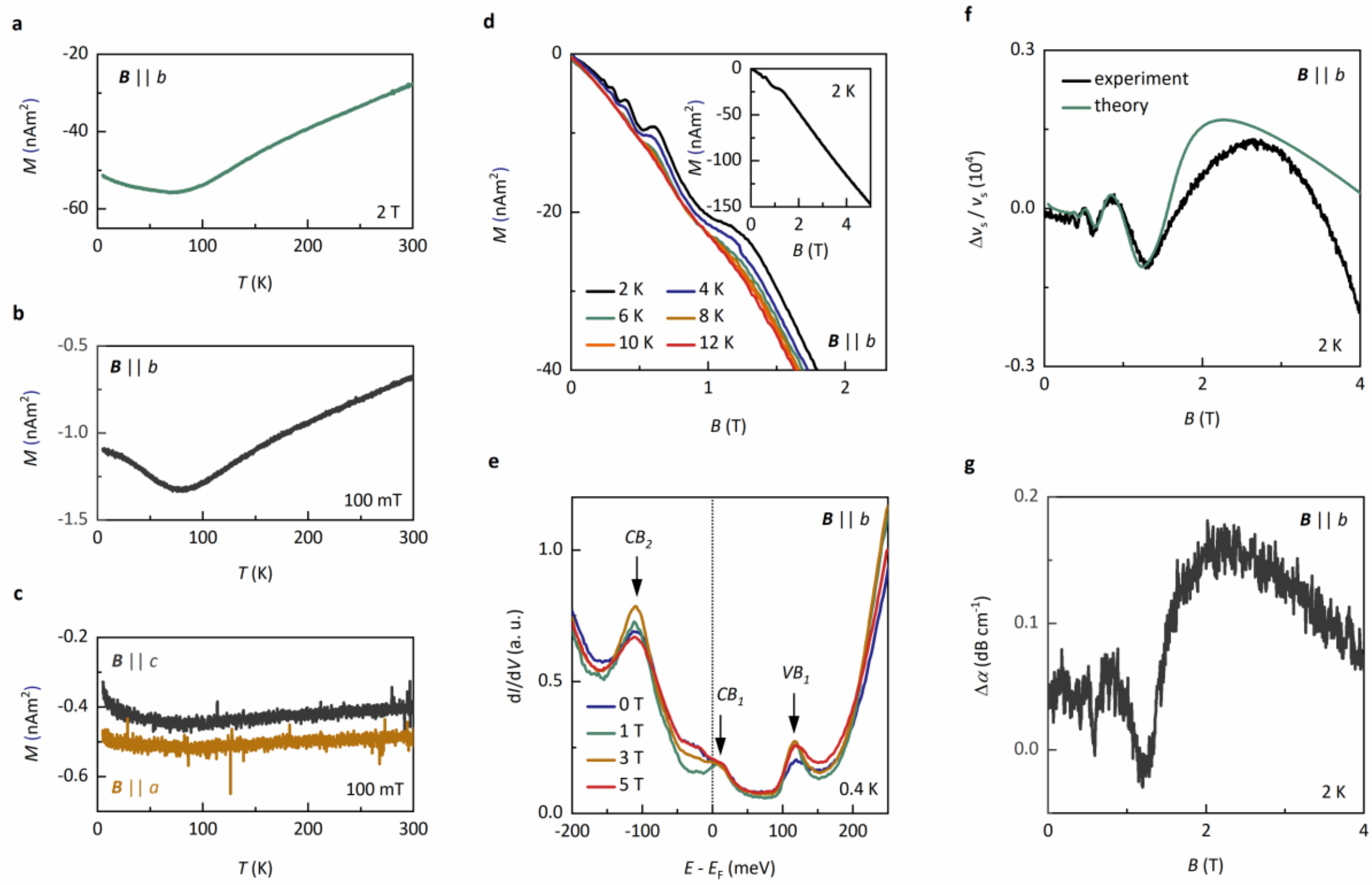

**Figure 2 | Probing the electronic spectrum of $ZrTe_5$ in magnetic fields. a**, Magnetization *M* as a function of temperature *T* with a 2 T magnetic field *B* applied along the *b*-axis of Sample C. **b**, *M* as a function of *T* with a 50 mT applied along the *b*-axis of Sample C. **c**, *M* as a function of *T* with a 100 mT applied along the *a*-axis (black line) and *c*-axis (dark yellow line) of Sample C. **d,** De Haas-van Alphen oscillations observed in *M* as a function of *B* applied along the *b*-axis at various *T* on Sample C. The inset shows a zoom-out of *M* as a function of *B* applied along the *b*-axis at 2 K. **e,** Differential conductance d*I*/d*V* as a function of bias voltage (energy *E*) with respect to the Fermi level $E_F$ (dotted line) for various *B* applied along the *b*-axis of Sample D at 0.4 K. The conduction band edges are marked with $CB_1$ and $CB_2$, the valence band edge is marked with $VB_1$. **f,** Sound velocity variation $\Delta v_s/v_s$ of the longitudinal mode (propagation along the *a*-axis, polarization vector along the *a*-axis) as a function of magnetic field applied along the *b*-axis at 2 K on Sample E. The black line is experimental data, the green line a theoretical model (see Methods), based on a Dirac Hamiltonian. **g**, Sound attenuation $\Delta\alpha$ of the longitudinal mode (propagation along the *a*-axis, polarization vector along the *a*-axis) as

a function of magnetic field applied along the *b*-axis at 2 K on Sample E. None of the experiments shows signatures of a charge-density-wave.

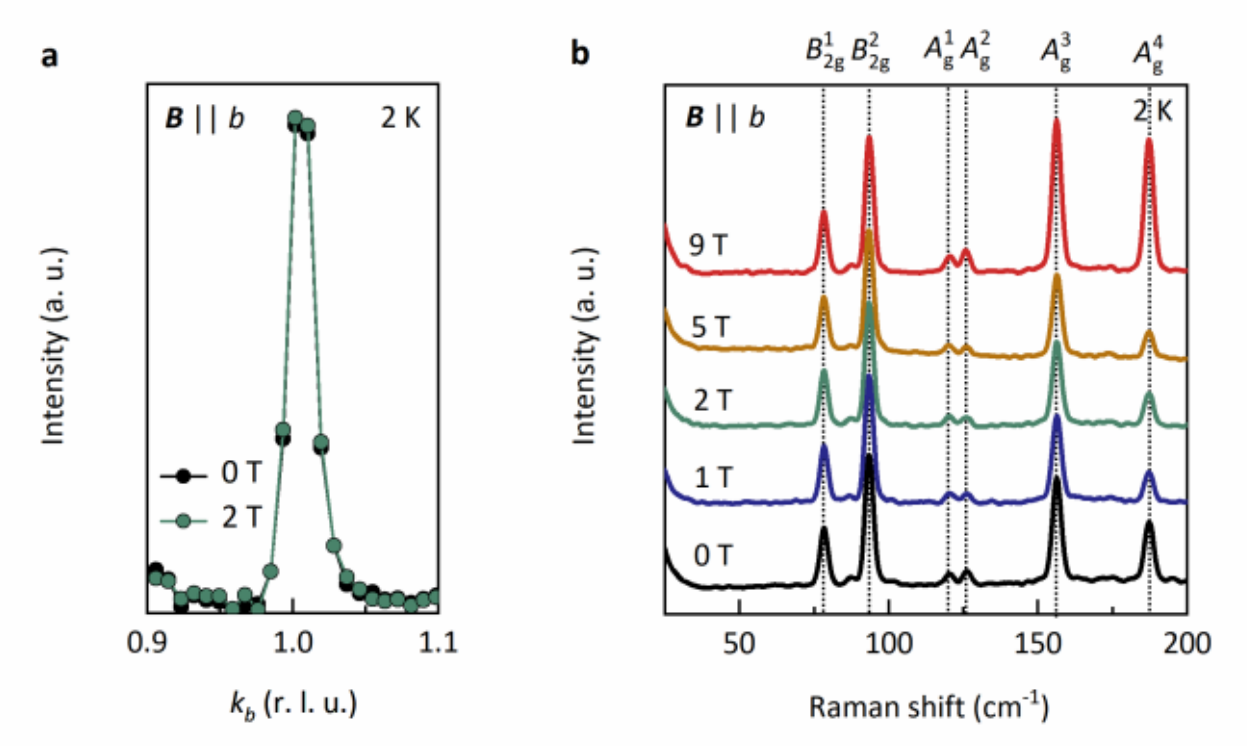

**Figure 3 | Probing the crystal lattice of $ZrTe_5$ in magnetic fields. a**, The (010)-peak observed in X-Ray diffraction on Sample A at magnetic fields $B$ of 0 T and 2 T at 2 K. **b**, Raman spectra of Sample F for various $B$ at 2 K. For fields up to 9 T, the Raman modes are located at 67, 84, 113, 118, 144, and 178 cm$^{-1}$, labelled with $B^1{}_{2g}$, $B^2{}_{2g}$, $A^1{}_g$, $A^2{}_g$, $A^3{}_g$, and $A^4{}_g$, respectively. None of the experiments shows signatures of a Charge Density Wave.

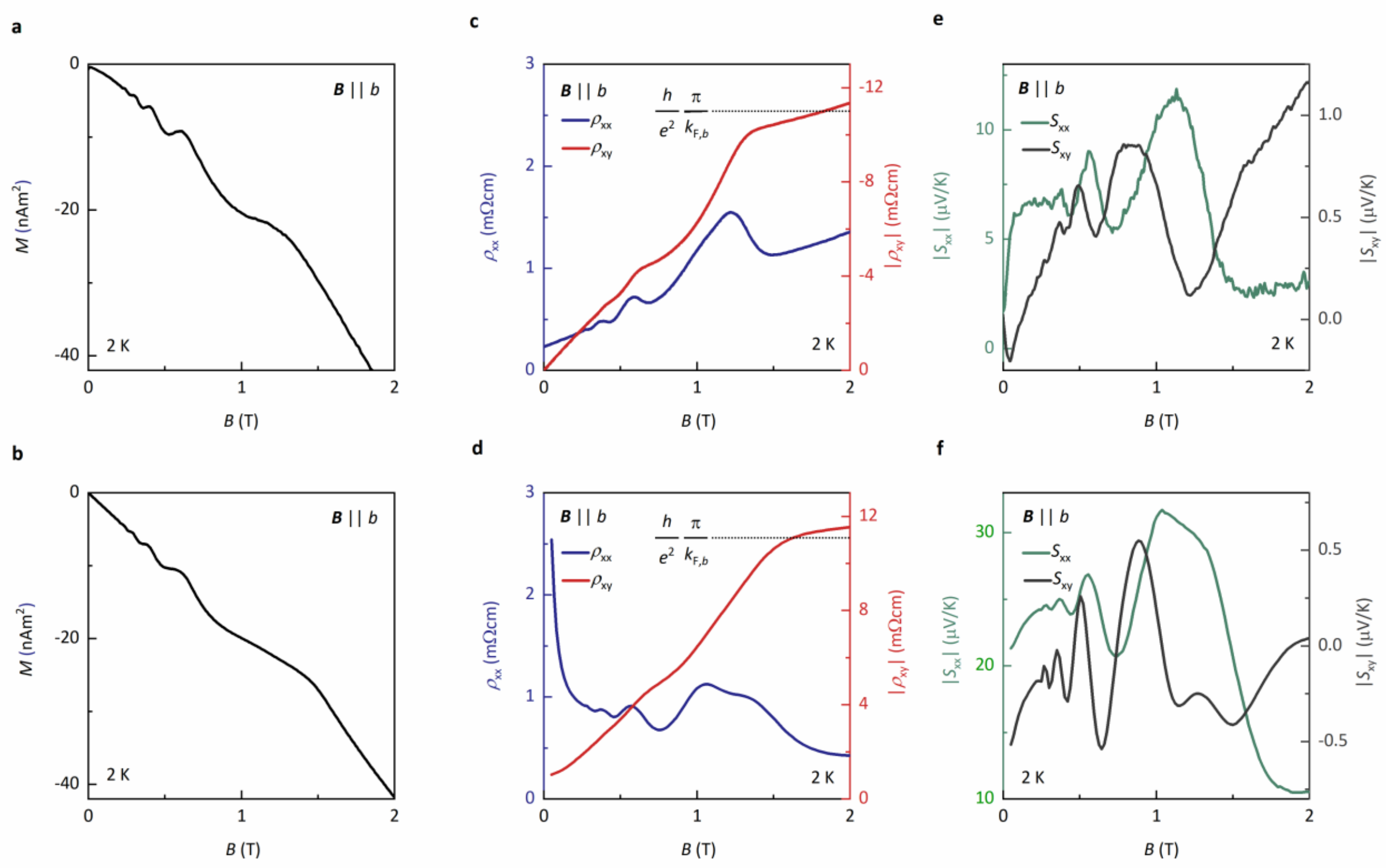

**Figure 4 | Comparison of the experimental and theoretical response of $ZrTe_5$ to magnetic fields *B*. a**, Measured and **b**, calculated magnetization *M* as a function of *B* applied along the *b*-axis of Sample G at 2 K. The magnetization plotted in **b,** is computed numerically from the derivative of the free energy. **c**, Measured and **d**, calculated longitudinal $\rho_{xx}$ (left axis) and Hall resistivity $\rho_{xy}$ (right axis) as a function of *B* applied along the *b*-axis of Sample A at 2 K. The calculations are done with a quantum lifetime $\tau_Q = 3.064$ ps at a temperature of 2 K. **e**, Measured and **f**, calculated Seebeck coefficient $S_{xx}$ and Nernst coefficient $S_{xy}$ as a function of *B* applied along the *b*-axis of Sample A at 2 K. The numerical data in **f** is calculated for the same parameters as the data shown in **d**. $\rho_{xx}$ and $\rho_{xy}$ in **d** and $S_{xx}$ and $S_{xy}$ in **f** are computed from linear response with a fixed chemical potential using the band structure parameters extracted from the experiments.